\documentclass{article}

\usepackage{arxiv}
\usepackage[section]{placeins}

\usepackage[utf8]{inputenc} 
\usepackage[T1]{fontenc}    
\usepackage{hyperref}       
\usepackage{url}            
\usepackage{booktabs}       
\usepackage{amsfonts}       
\usepackage{nicefrac}       
\usepackage{microtype}      
\usepackage{lipsum}		
\usepackage{graphicx}
\usepackage{natbib}
\usepackage{doi}

\title{Enhancing Autoignition Characteristics: A Framework to Discover Fuel Additives and Making Predictions Using Machine Learning}

\author{ {Shahid Rabbani} \\
	Mechanical Engineering Department\\
	Khalifa University\\
	Abu Dhabi, UAE \\
	{shahid.rabbani@ku.ac.ae}\\
}

\renewcommand{\shorttitle}{\textit{arXiv}}

\hypersetup{
pdftitle={Enhancing Autoignition Characteristics: A Framework to Discover Fuel Additives and Making Predictions Using Machine Learning},
pdfsubject={q-bio.NC, q-bio.QM},
pdfauthor={Shahid Rabbani},
pdfkeywords={First keyword, Second keyword, More},
}

\begin{document}
\maketitle

\begin{abstract}
	Combustion process can become more energy efficient and environment friendly if used with appropriate fuel additive. Discovery of fuel additive can be accelerated by applying hybrid approach of using of chemical kinetics and Machine Learning (ML). In this work, we present a framework that takes the robustness of Machine Learning and accuracy of chemical kinetics to predict the effect of fuel additive on autoignition process. We present a case of making predictions for Ignition Delay Time (IDT) of biofuel n-butanol ($C_4H_9OH$) with several fuel additives. The proposed framework was able to predict IDT of autoignition with high accuracy when used with unseen additives. This framework highlights the potential of ML to exploit chemical mechanisms in exploring and developing the fuel additives to obtain the desirable autoignition characteristics.
\end{abstract}


\section{Introduction}
	In the wake of fast changing global climate scenario, a lot of emphasis has been laid on reducing hazardous emissions and using renewable energy solutions. Specially during the last few years, climate change has become a global challenge and many regions have already started experiencing its impact. Fortunately, during the same time period, ML algorithms have come of age and helping researchers in tackling variety of challenges. Also for climate change, ML is proving to be very helpful in suggesting ways to reduce emissions \cite{rolnick2019tackling}. For conventional fuels - which have been found to contribute significantly to increase emissions - ML has high potential to find ways to reduce emissions. For example, Li et al. used ML approach to explore organic waste to find equivalent renewable energy source of fossil fuels \cite{li2020fuel}. Badra et al. used combined approach of Computational Fluid Dynamics and ML to optimize combustion process \cite{badra2020combustion}.  Despite all such applications of ML to minimize effects of climate change and making combustion process more environment friendly, it can be noted that not much attention has been devoted to find fuel additives which can provide desirable emissions related characteristics of burning fuels.

In this work, we present a framework that uses ML algorithm and chemical kinetics to discover fuel additives. First we present the methodology that was employed to obtain data and train the model, and then we present results for fuel additives obtained using the framework.

\section{Methododology}
\label{sec:headings}
As an application of the framework, we present a case of finding IDT for autoignition of n-butanol. The approach of exploring new additives using this framework consist of three main steps:
 first step consists of obtaining results of IDT for n-butanol using experimentally validated chemical kinetics mechanism \cite{black2010bio}  which consists of 6 elements (C, H, N, O, Ar and He), 243 species and 2892 unidirectional reactions. Adiabatic autoignition of butanol at constant-volume was considered for simulations. Apart from n-butanol, total of 50 stable species were found in the mechanism which were used as additive in volumetric ratios of 0.0 (pure n-butanol), 0.01, 0.1, 0.2, 0.4, 0.6 and 0.8. IDT were obtained by running separate simulations for all these additives. For each additive, above mentioned volumetric ratios were considered in combination of different input conditions of temperature, pressure and stoichiometric ratios. Details of these input parameters are given in Table \ref{tab:parameters}. 
 
 \begin{table}[t]
\caption{Range of parameters used to generate data set}
\begin{center}
\begin{tabular}{ll}
\hline
\textbf{Input}                      & \textbf{Value}   \\ \hline
Temperature (K)      & 1100, 1350, 1600, 1800  \\
Pressure (atm)        & 1, 4, 8   \\
Equivalence Ratios                & 0.6, 1.0, 1.8  \\
Number of Additives        & 50 \\
Additives Mixing Ratios (vol.) & 0.0, 0.01, 0.1, 0.2, 0.4, 0.6, 0.8   \\ \hline
\end{tabular}
\label{tab:parameters}
\end{center}
\end{table}

 It should be noted that all these combinations of initial conditions do not necessarily lead to ignition, therefore those simulations that did not result in ignition were omitted from the data such that total number of IDT data points obtained in this work count to 11,732.
 
\begin{figure}[!htbp]]
\centering
\includegraphics[scale=0.4]{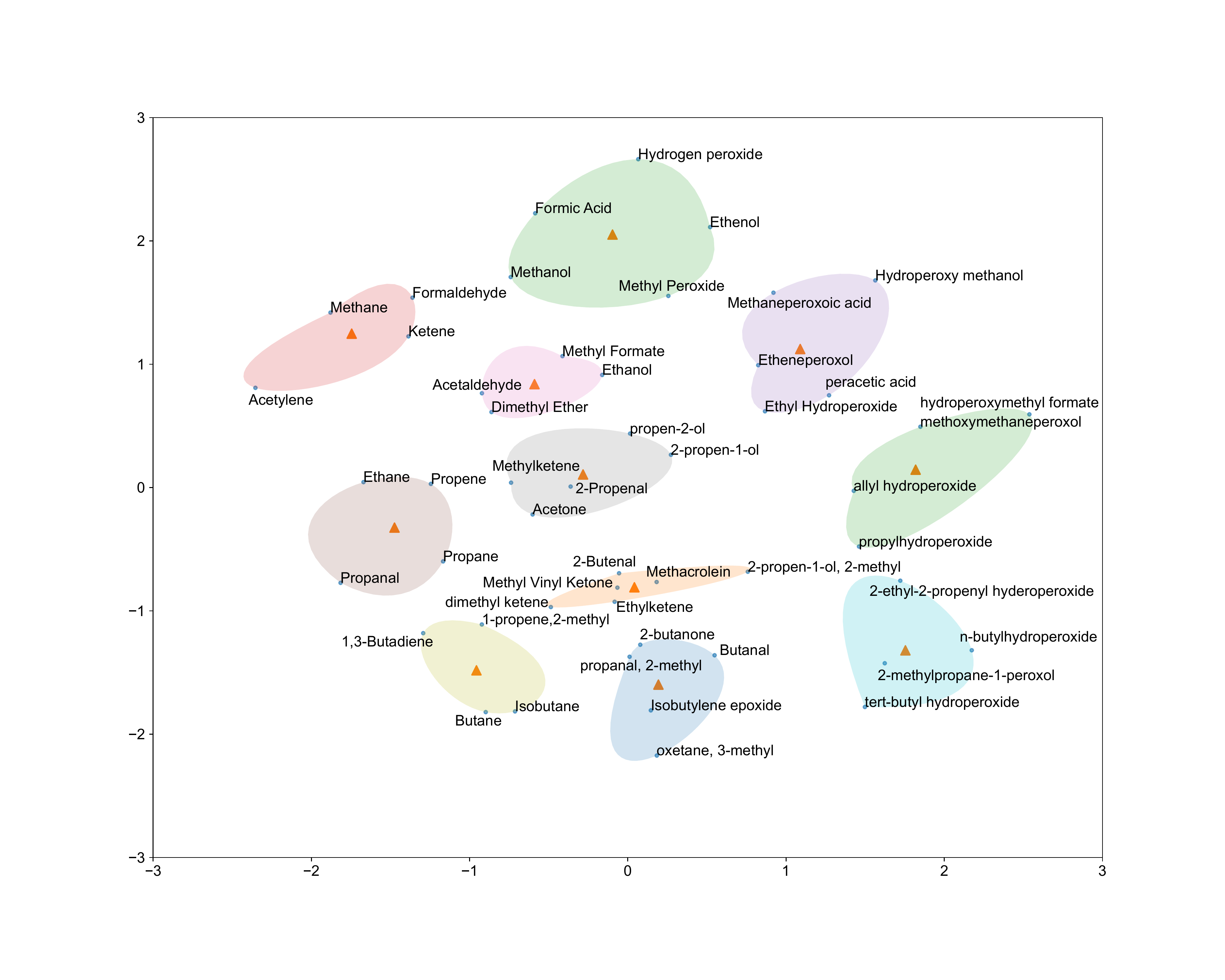}
\caption{Clusters of fuel additives presented in non-dimensional space with respect to their features. Blue circles indicate spatial position while orange triangles denote centroid of  clusters.} 
\label{fig:Cluster}
\end{figure}
 
 Second step comprises of finding the features of all 50 additives used in step 1 to characterize the fuel additive. In this work, we divided features into five categories: (i) Chemical (3D atom count, 3D anion-count etc.), (ii) Physical (Total Polar Surface Area (TPSA), volume of molecule, XlogP etc.), (iii) Compositional (no. of C atoms, molecular weight etc.) and (iv) Structural (bag of bonds such as no. of O-O bonds, number of C=C bonds etc.) and (v) Thermodynamic properties (coefficients of polynomials used to represent thermodynamic data used in NASA chemical equilibrium code \cite{gordon1994computer} etc.). All these features sum up to 46. Figure \ref{fig:Cluster} shows additives plotted using Multi-Dimensional Scaling (MDS) of all features such that similar additives cluster together. For example, butane and iso-butane have many features in common, hence they share close proximity.
 
\begin{figure}[!htbp]]
\centering
{\includegraphics[scale=0.195]{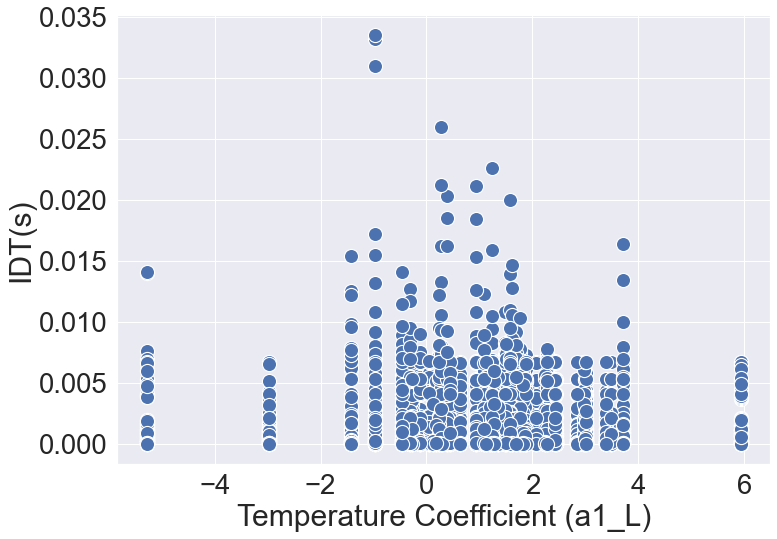}}  \includegraphics[scale=0.195]{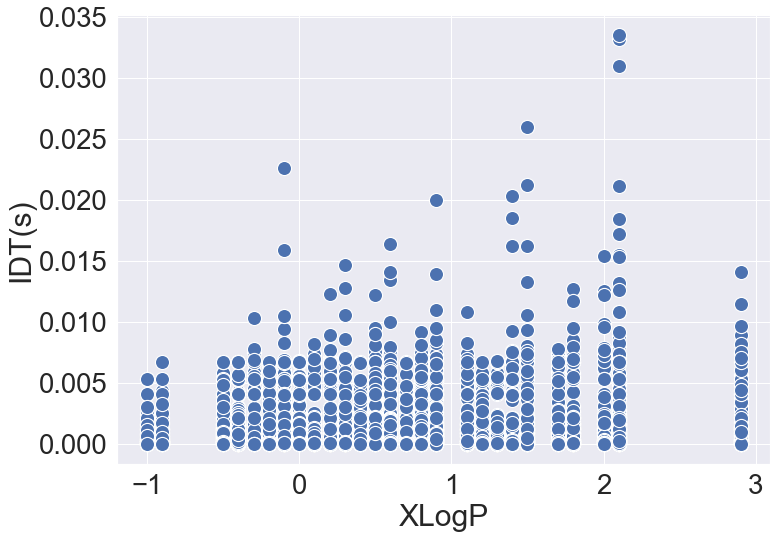}  \includegraphics[scale=0.195]{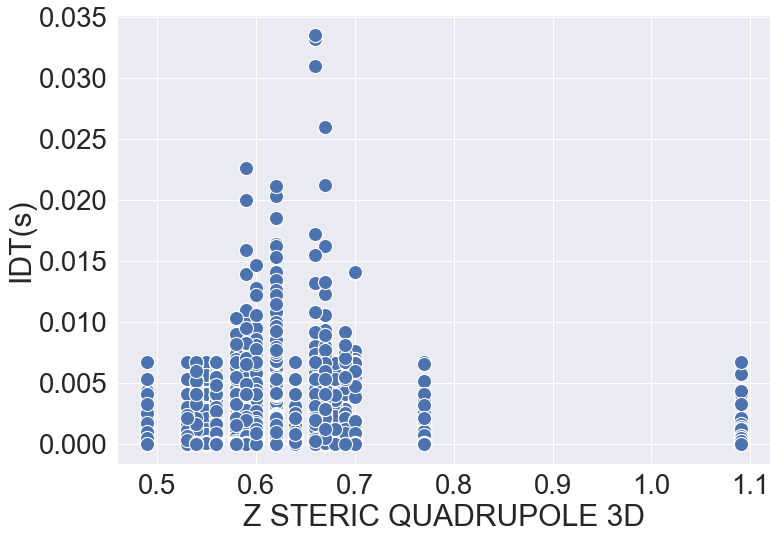} \\
{\includegraphics[scale=0.195]{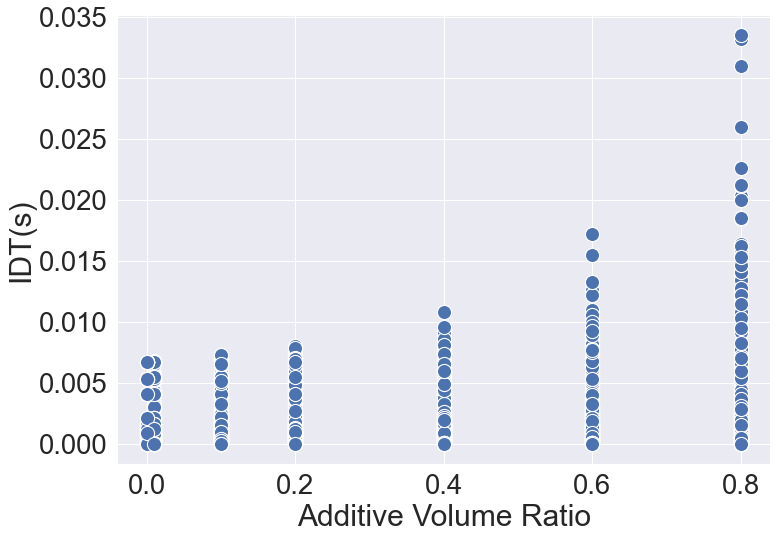} }  \includegraphics[scale=0.195]{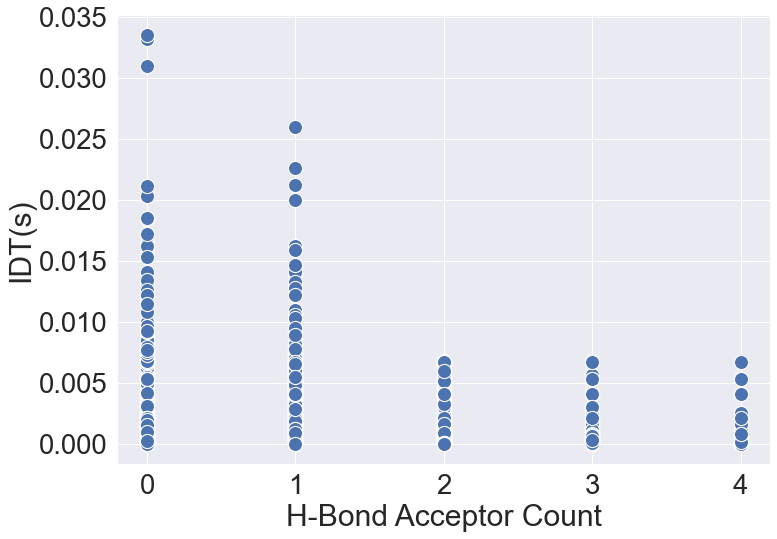} \hspace{3 pt} \includegraphics[scale=0.195]{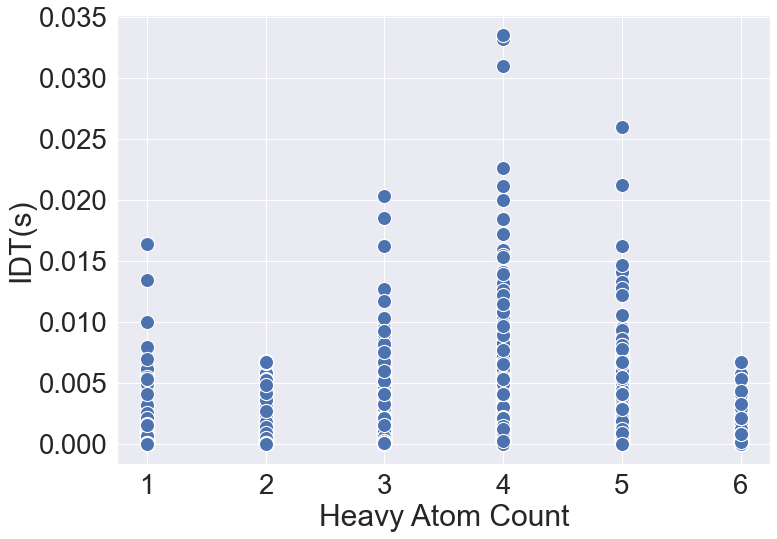}
\caption{Scatter plots of IDT with few features of fuel additives.}
\label{fig:Parameter_scatter}
\end{figure}

Figure \ref{fig:Parameter_scatter} shows distribution of IDT obtained in step 1 when plotted against features obtained in step 2. Although in this figure, IDT is plotted against only six features, yet it can be seen that features of fuel additives relate to IDT with clear patterns. Third step is to exploit such patterns of additive features with IDT using ML. Deep Neural Network (DNN) were employed in this work to fit the IDT with additives features and initial conditions. First a DNN model was generated with full data of all 50 additives such that 80\% of data was used to train the model while 20\% data was used for testing purpose. This model was tested to predict the IDT with the different initial conditions of additives for which model was trained. 

Once it was established that model can predict the IDT with additives which were part of DNN training, the study was extended to predict the IDT for additives which were not trained in the network so that the capability of this framework to predict the effect of new additives can be assessed. To achieve this objective, another DNN model was trained with only 48 additives and results were tested for two additives which were not part of new trained model. 
In the next sections, first the results of trained model for 50 species are presented, followed by the results of DNN model trained on 48 additives to predict IDT for two unseen additives. Here unseen is referred to the additives which were not part of DNN trained on 48 additives.

\section{Results}

This section is divided into two parts. In the first part results of the multi-layer DNN are presented where model was trained and tested for 50 additives. The second part relates to the DNN model which was trained for 48 additives and was tested for two unseen additives.

\subsection{Evaluation for all 50 additives}

\begin{figure}[t]
\centering
\includegraphics[scale=0.4]{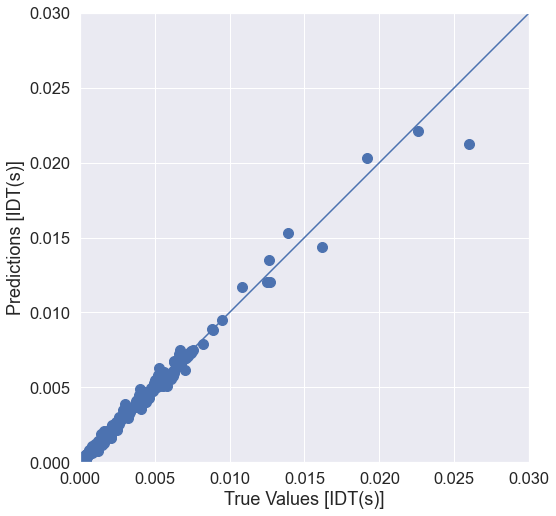}
\caption{Comparison of predicted vs true values of IDT for test data using the model trained on  50 additives.} 
\label{fig:all_50}
\end{figure}

\begin{figure}[t]
\centering
\includegraphics[scale=0.4]{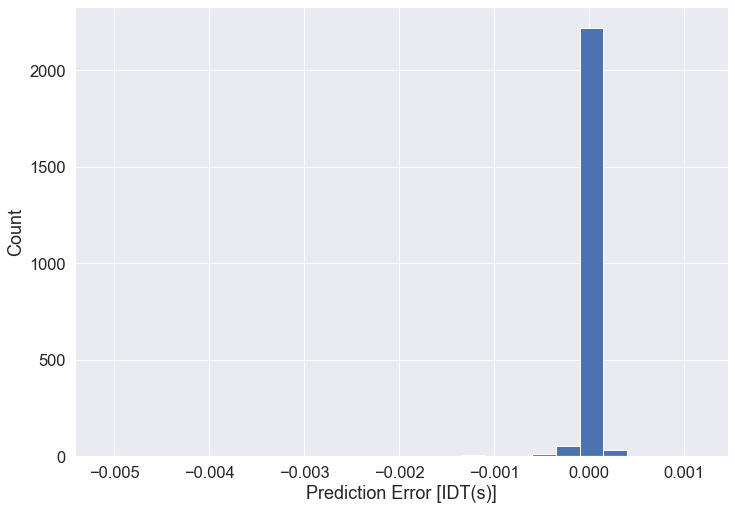}
\caption{Error distribution for predictions of test data using model trained on 50 additives.} 
\label{fig:error}
\end{figure}

Figure \ref{fig:all_50} shows the comparison between true values of IDT – which were obtained from autoignition simulations - with IDT obtained from DNN model trained on 50 species. This figure shows the predictions against randomly selected test data points which are 20\% of all the available data for 50 species.
Overall R2 score for the test data was 0.99. It can be seen that most of the IDT values are below 0.10 s and less than 10 values are above 0.10 s which have relatively high error. This distribution of error can also be seen in Figure \ref{fig:error}. As can be seen, this distribution is reasonable because most of the data used for training has IDTs below 0.10 s which leads to high accuracy for region below 0.10 s. 

\subsection{Evaluations for unseen additives}
To test the framework for unseen additives, data points for two additives were completely omitted from the training data and a new DNN was trained on 48 additives. The two unseen additives were ethane ($C_2H_6$) and methyl-vinyl-ketone ($C_2H_3COCH_3$). It can be seen from Figure \ref{fig:Cluster} that $C_2H_3COCH_3$ has features which are similar to several other neighboring fuel additives such as ethyl ketene and 2-butenal. This shows that although $C_2H_3COCH_3$ is not included in the training data, however training data contains fuel additives which have similar features. Unlike $C_2H_3COCH_3$, $C_2H_6$ does not share close proximity with other fuel additives. This indicates that training data does not contain fuel additives which have as similar features to $C_2H_6$ as $C_2H_3COCH_3$. Using this selection of additives will enable to assess the capability of model to predict IDT for unseen additives; irrespective of similarity in features with training data.

\begin{figure}[]
\centering
\includegraphics[scale=0.4]{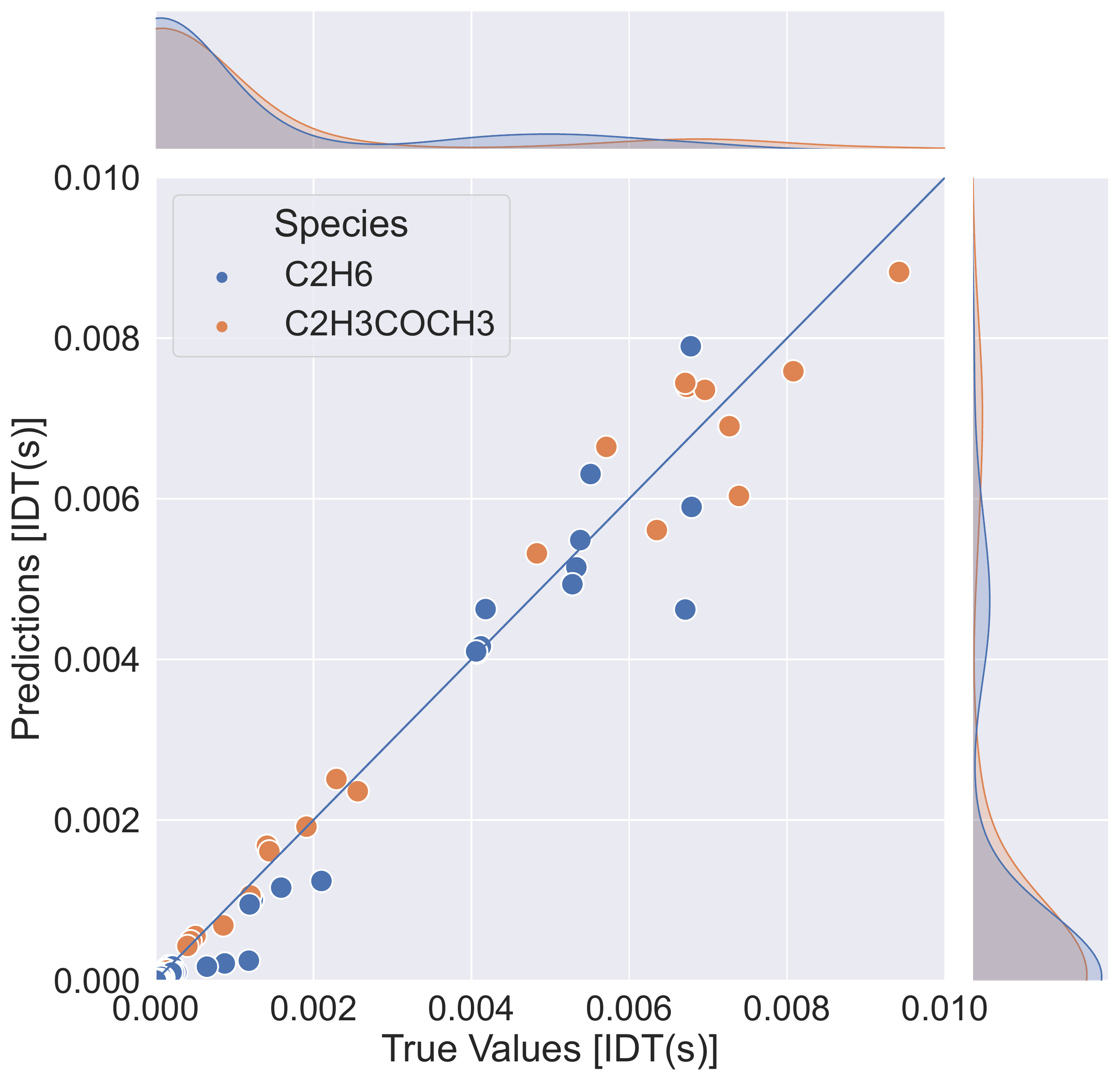}
\caption{Comparison of predicted vs true values of IDT for unseen additives using model trained on  48 additives. Top and right plots show distribution of data points.} 
\label{fig:Unseen}
\end{figure}

Figure \ref{fig:Unseen} shows that result of IDT predictions against true values for $C_2H_6$ and $C_2H_3COCH_3$. It can be seen that predicted values are close to true IDT values. Similar to Figure \ref{fig:all_50}, most of the data points locate below 0.004 s. Although the points are more dispersed as compared to DNN trained on 50 species, however R2 score is still 0.97 which indicates high accuracy. So, the trained model on 48 additives successfully able to predict IDT for the unseen additives.
\vspace {0mm}
\section{Conclusions}

In this work, a framework to predict autoignition characteristics -both for seen and unseen additives - is presented. The framework combines the accuracy of experimentally validated chemical mechanism and robustness of ML to predict autoignition characteristics. An example of renewable fuel n-butanol is presented to predict IDT for trained and untrained fuel additives. It was shown that the framework was able to capture the chemical kinetics to predict IDT for the additives included in the chemical mechanism. Moreover, framework was also successfully able to predict IDT for species which were not part of DNN trained model. As shown using the case of unseen additives, this work also highlights the applicability of this framework to study fuel additives which that are not part of chemical kinetic mechanism, thus opening a whole new domain to explore and discover new fuel additives to achieve desired autoignition characteristics. In summary, the framework can be used to:

\begin{itemize}
\item Study the effect of new fuel additives; irrespective of their presence in chemical kinetics mechanism. 
\item Predict the effect of new additives on emission such as $NO_x$, $CO$, $CO_2$, $CH_2O$ etc.
\item Predict maximum heat generation during autoignition process.
\item Predict adiabatic flame temperature during autoignition process. 
\item Predict flame-type using new additive.
\end{itemize}

\bibliographystyle{unsrtnat}
\bibliography{references} 

\begin{thebibliography}{5}
\providecommand{\natexlab}[1]{#1}
\providecommand{\url}[1]{\texttt{#1}}
\expandafter\ifx\csname urlstyle\endcsname\relax
  \providecommand{\doi}[1]{doi: #1}\else
  \providecommand{\doi}{doi: \begingroup \urlstyle{rm}\Url}\fi

\bibitem[Rolnick et~al.(2019)Rolnick, Donti, Kaack, Kochanski, Lacoste,
  Sankaran, Ross, Milojevic-Dupont, Jaques, Waldman-Brown,
  et~al.]{rolnick2019tackling}
David Rolnick, Priya~L Donti, Lynn~H Kaack, Kelly Kochanski, Alexandre Lacoste,
  Kris Sankaran, Andrew~Slavin Ross, Nikola Milojevic-Dupont, Natasha Jaques,
  Anna Waldman-Brown, et~al.
\newblock Tackling climate change with machine learning.
\newblock \emph{arXiv preprint arXiv:1906.05433}, 2019.

\bibitem[Li et~al.(2020)Li, Pan, Suvarna, Tong, and Wang]{li2020fuel}
Jie Li, Lanjia Pan, Manu Suvarna, Yen~Wah Tong, and Xiaonan Wang.
\newblock Fuel properties of hydrochar and pyrochar: Prediction and exploration
  with machine learning.
\newblock \emph{Applied Energy}, 269:\penalty0 115166, 2020.

\bibitem[Badra et~al.(2020)Badra, Sim, Pei, Viollet, Pal, Futterer, Brenner,
  Som, Farooq, Chang, et~al.]{badra2020combustion}
Jihad Badra, Jaeheon Sim, Yuanjiang Pei, Yoann Viollet, Pinaki Pal, Carsten
  Futterer, Mattia Brenner, Sibendu Som, Aamir Farooq, Junseok Chang, et~al.
\newblock Combustion system optimization of a light-duty gci engine using cfd
  and machine learning.
\newblock Technical report, SAE Technical Paper, 2020.

\bibitem[Black et~al.(2010)Black, Curran, Pichon, Simmie, and
  Zhukov]{black2010bio}
G~Black, HJ~Curran, S~Pichon, JM~Simmie, and V~Zhukov.
\newblock Bio-butanol: Combustion properties and detailed chemical kinetic
  model.
\newblock \emph{Combustion and Flame}, 157\penalty0 (2):\penalty0 363--373,
  2010.

\bibitem[Gordon and McBride(1994)]{gordon1994computer}
Sanford Gordon and Bonnie~J McBride.
\newblock Computer program for calculation of complex chemical equilibrium.
\newblock \emph{NASA reference publication}, 1311, 1994.

\end{thebibliography}







\end{document}